\documentclass[12pt]{iopart}

\usepackage{cite}
\usepackage{hyperref}
\hypersetup{colorlinks=true, citecolor=blue, urlcolor=blue, linkcolor=blue}
\usepackage[english]{babel}
\usepackage[utf8]{inputenc}
\usepackage{graphicx}
\usepackage{color,soul}

\usepackage{amsfonts}
\usepackage{amssymb}
\usepackage{iopams} 
\usepackage{caption}
\begin{document}

\title{Reinforcement learning of optimal active particle navigation}

\author{Mahdi Nasiri $^1$ and Benno Liebchen$^1$}

\address{$^1$ Institute of Condensed Matter Physics, Technische Universität Darmstadt,\\
\hspace{0.15 cm} D-64289 Darmstadt, Germany.}

\ead{benno.liebchen@pkm.tu-darmstadt.de}

\vspace{10pt}
\begin{indented}
\item[]\today
\end{indented}

\begin{abstract}
The development of self-propelled particles at the micro- and the nanoscale has sparked a huge potential for future applications in active matter physics, microsurgery, and targeted drug delivery. However, while the latter applications provoke the quest on how to optimally navigate towards a target, such as e.g. a cancer cell, there is still no simple way known to determine the optimal route in sufficiently complex environments. Here we develop a machine learning-based approach that allows us, for the first time, to determine the asymptotically optimal path of a self-propelled agent which can freely steer in complex environments. Our method hinges on policy gradient-based deep reinforcement learning techniques and, crucially, does not require any reward shaping or heuristics. The presented method provides a powerful alternative to current analytical methods to calculate optimal trajectories and opens a route towards a universal path planner for future intelligent active particles.
\end{abstract}

%
\vspace{2pc}

%
%
%

\section*{Introduction}
The problem of finding suitable navigation strategies 
is of great interest to applications ranging from motion planning for autonomous underwater vehicles, ocean gliders \cite{hwan2011anytime, petres2007path, panda2020comprehensive, zeng2016comparison}, and aerial vehicles \cite{ChitsazAir, techy2009minimum, guerrero2013uav} to microorganisms searching for food and prey \cite{fricke2016persistence,fricke2016immune} and striving for survival in complex environments \cite{volpe2017topography,ipina2019bacteria}.
One important class of path planning problems which is currently attracting a rapidly increasing attention is centered around the quest for the optimal trajectory allowing an active particle, which can freely steer but cannot control its speed, to 
reach a given target in a complex environment. This active particle navigation (APN) problem is relevant both for biological swimmers like fish or for turtles on the way to their breeding grounds 
\cite{hays2014route,mclaren2014optimal} and for future applications of synthetic microswimmers \cite{li2017micro} such as targeted drug \cite{erkoc2019mobile, yasa2018microalga,luo2018micro} and gene delivery \cite{esteban2016acoustically,hansen2018active} or microsurgery \cite{vyskocil2020cancer}. \\
Contrasting classical navigation problems of vehicles like ships or airplanes \cite{https://doi.org/10.1002/zamm.19310110205,5531602}, mesoscopic active particles can face a variety of new challenges including fluctuations, hydrodynamic interactions with obstacles and boundaries \cite{daddi2021hydrodynamics, PhysRevFluids.5.082101}, and highly complex environments \cite{volpe2011microswimmers, spagnolie2015geometric,PhysRevLett.118.158004, PhysRevLett.116.028104}.
Following these complex ingredients, the optimal (fastest) path generically differs from the shortest one and is highly challenging to determine.
In fact, as of now, there is no standard receipe to find the optimal path even for the simplest case of a dry active particle with perfect steering (no fluctuations, no delay) \cite{liebchen2019optimal} in \textit{sufficiently complex environments}. While a method to tackle this challenge, which we refer to as the "minimal APN problem", would serve as a crucial step forward en route to the ultimate dream of a universal path planner for microswimmers, 
such a method is not yet available. \\
In fact, classical path planning algorithms such as A* or Dijkstra can not be straightforwardly adjusted to account for complex smooth environments (even when using heuristic ingredients) and achieving general solutions with analytical methods based on variational methods \cite{liebchen2019optimal}, 
optimal control theory \cite{kirk2004optimal, daddi2021hydrodynamics} or transformations 
to a problem of finding geodesics of a Randers metric \cite{piro2020optimal}
is highly challenging (if not impossible) if the environment is sufficiently complex. In contrast, recent advancements in the field of 
artificial intelligence and reinforcement learning (RL) have allowed handling the required complexity and can 
be used in principle to approach the (minimal) APN problem for generic environments. 
Corresponding pioneering studies at the interface of RL and active matter 
\cite{cichos2020machine} have demonstrated, remarkably, that 
tabular Q-learning algorithms are able to uncover efficient navigation strategies for certain environments \cite{schneider2019optimal,muinos2021reinforcement} including complex fluid flows \cite{PhysRevLett.118.158004}. Other RL based methods 
have been used to develop efficient navigation strategies even in the presence of turbulent and chaotic flows \cite{gustavsson2017finding, biferale2019zermelo} or learning and modeling chemotaxis behavior \cite{Hartle2019683118}. Deep RL methods 
have been applied to 
colloidal robots in unknown environment with random obstacles \cite{yang2020efficient, yang2020micro} and very recently Q-learning was used to explore optimal colloidal predator-prey dynamics and strategies \cite{PhysRevE.104.054614}.\\
However, despite these successes, the challenge of finding the globally optimal path in generic environments still remains. First, there is generally a major risk that agents converge to a local optimum rather than to the global one \cite{sutton2018reinforcement}. Second, even after finding the global optimum 
for given reward and state-action space definitions, it often remains unclear how 
the result 
relates to optimality in the physical reality. In some cases this deviation can reach a level where the 
final result is nowhere near the optimal trajectory, as we'll see below.
\\To overcome this gap in the literature between methods which can not handle the complexity of generic complex environments and methods leading to solutions which do not (or are not known to) lead to the globally optimal path, in the present work we develop a new approach 
which does not only asymptotically reproduce analytically known optimal solutions for the minimal APN problem but also allows to find the optimal path in very complex environments.\\
To achieve this, we combine a hybrid discrete-continuum representation of the environment with policy gradient based Deep RL agents (advantageous actor-critic algorithms) to find the optimal path.

\begin{figure}[t] 
\includegraphics[scale=0.4]{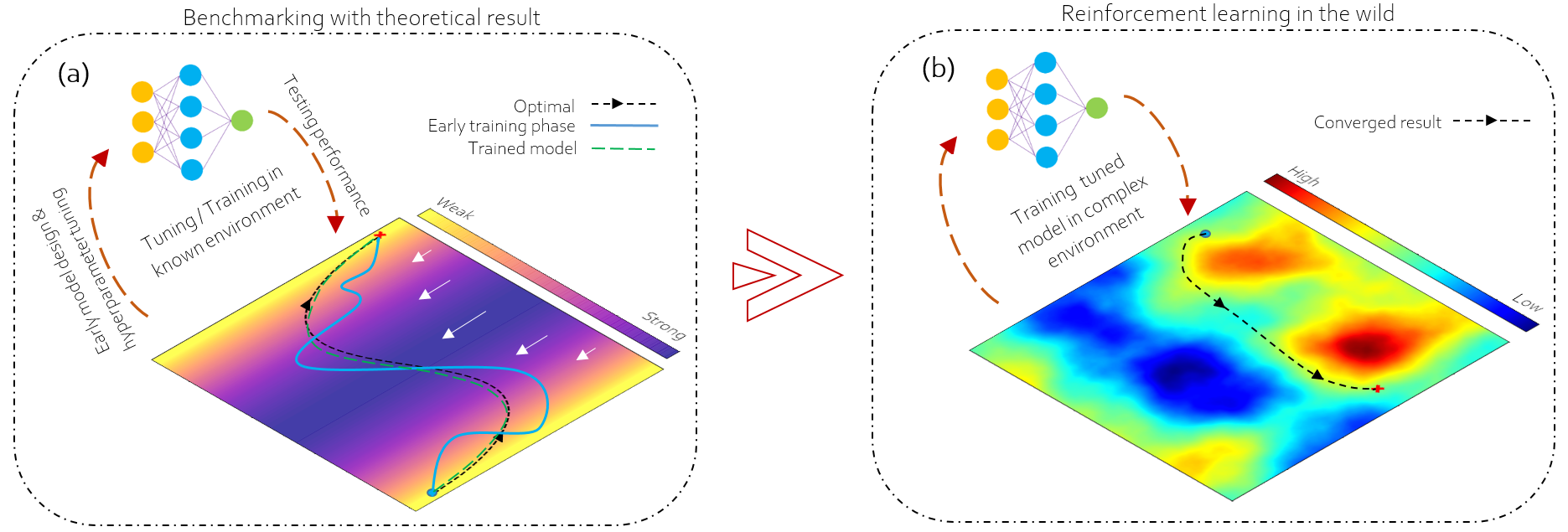}
\caption{Schematic illustration of the developed machine learning approach.
(a) The RL model is designed, trained and fine-tuned to find the optimal trajectory in environments with known analytical solutions. 
Upon training, the model's early performance (blue curve) converges asymptotically (green dashed curve) to the optimal solution (dashed black curve). 
Arrows and colors indicate the direction and strength of force/flow fields acting on the agent. 
(b) The designed model with proven performance is trained 
in 
highly complex environments to provide asymptotically optimal solutions in cases which are analytically inaccessible. 
The dashed line exemplarically shows the learned path in a Gaussian random potential (background colors). 
}
\label{fig:Novelty}
\end{figure} 

\noindent Remarkably, we find that using a single relatively simple neural network architecture across different environments, is sufficient to achieve asymptotically optimal trajectories i.e. trajectories which are optimal for a given discretization. Our results unify, for the first time, asymptotic optimality with the feasibility of handling generic complex   environments. They can also be interpreted as an indication that there is a way for biological agents to learn \emph{optimal} navigation strategies from generation to generation (e.g. throughout evolution) without requiring direct global knowledge of their environment or even of the location of their target.

\section*{Model}
To define the minimal APN problem, we consider an overdamped dry active particle in 2D at position $\vec{r}(t) = (x(t), y(t))$ which evolves as:
\begin{eqnarray}
    \dot{\vec{r}}(t) = v_{0}\hat{n}(t)+\vec{f}(\vec{r}) 
    \label{eq1}
\end{eqnarray}

\noindent Here, $v_0$ is the constant self-propulsion velocity of the active particle
and $\vec f(\vec r)=\vec{F}(\vec{r})/\gamma(\vec r)+\vec u(\vec r)$ 
is the overall external field with $\vec F(\vec r),\vec u(\vec r)$ being the force field and the solvent flow field due to the environment and $\gamma(\vec r)$ is the 
Stokes drag coefficient which can vary in space as relevant e.g. for viscotaxis \cite{liebchen2018viscotaxis}. 
We assume that the steering direction $\hat{n}(t) = (\cos \psi(t), \sin \psi(t))$ can be freely controlled by the active particle, as relevant e.g. for biological microswimmers, 
or by external fields e.g. via feedback control systems \cite{khadka2018active,lavergne2019group,sprenger2020active} or external electric, magnetic or phoretic fields \cite{demirors2018active,liebchen2018synthetic,stark2018artificial}.\\
While in reality there would of course be some delay in the control \cite{khadem2019delayed} as well as thermal fluctuations (for small agents), here we 
neglect both complications 
as even in their absence there is no known way to 
systematically determine or learn globally optimal trajectories in the literature. 
However, we would like to stress that our approach can be generalized to include both ingredients. 
To complete the definition of the minimal APN problem, let us now assume that the starting $\vec{r}(0) = \vec{r}_{start}$ and target $\vec{r}(T) = \vec{r}_{end}$ points as well as $\vec{f}(\vec{r})$ are given and the task is to find the connecting path (and the associated navigation strategy) which minimizes the travelling time.

\section*{Approach}

\begin{figure}[t] 
\includegraphics[scale=0.335]{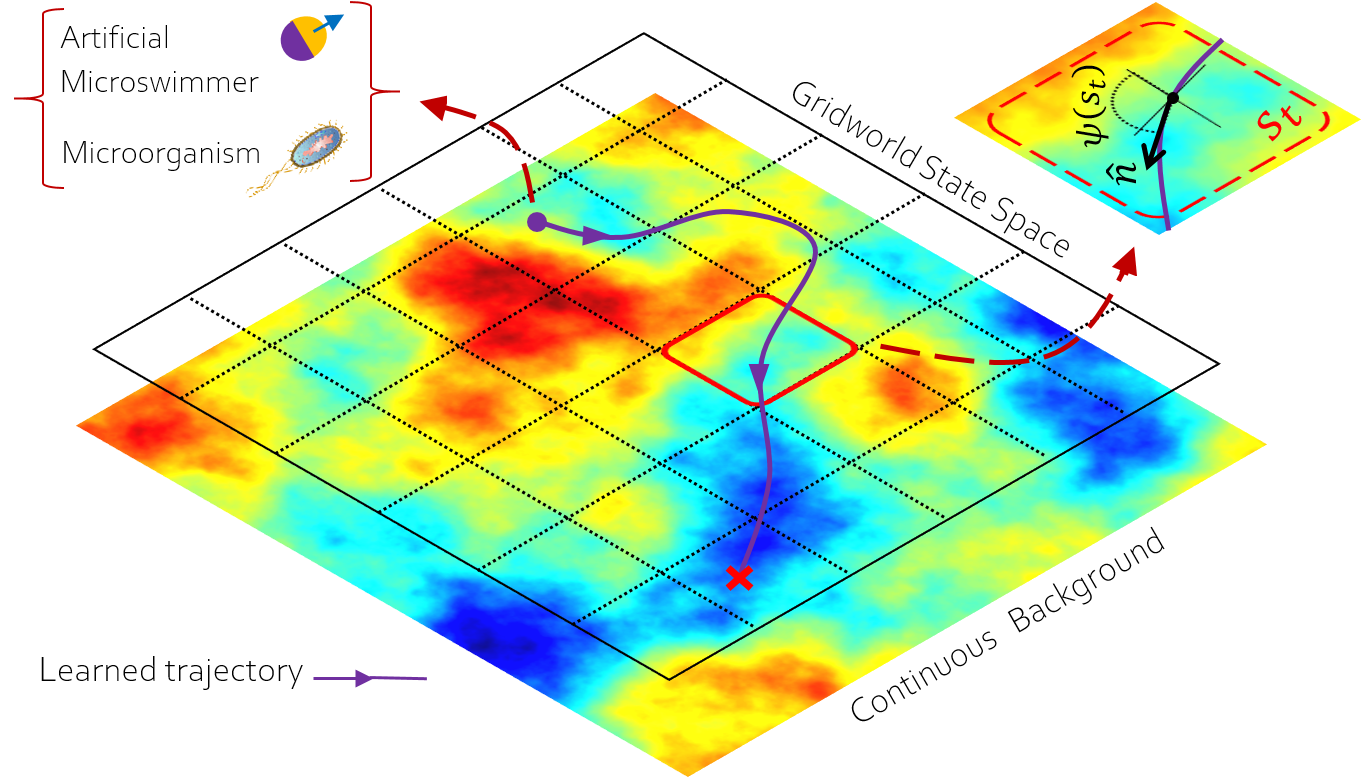}
\centering
\caption{Illustration of the 
hybrid combination of continuous and discrete state spaces used by our approach. The actions and effects of the force fields take place in a continuous environment and only a discrete observation is fed back into the model.}
\label{fig:Hybrid}
\end{figure}

Here we use reinforcement learning to solve the minimal APN problem. We train an agent {within} a complex environment, to take actions in order to maximize a cumulative reward \cite{kaelbling1996reinforcement,sutton2018reinforcement}. 
Within each episode, the 
agent starts from a certain initial state $s_0$ (initial position), and chooses an action $a_t$ (direction of motion) at each time step $t$ until it either reaches the target point (or region) or 
faces an exiting condition, e.g. by hitting the boundaries of
the simulation domain or exceeding the maximum number of allowed steps (see SI for details). 
\\For this study, we consider a hybrid interplay of discrete and continuous state spaces for the environment so that the RL agent can harness the power of working inside an infinite dimensional continuous state space without dealing with high dimensional input data (Fig. \ref{fig:Hybrid}), which is crucial for the present approach. 
That is, we account for the exact continuous external field $\vec f(\vec r)$, but discretize the environment, represented as a 2D gridworld observation, when feeding it 
into the neural networks as the input states, $s_t$.
In each step, the agent chooses an action from a set of 
60 equally spaced orientation angles $\psi(s_t) = \{ \frac{m\pi}{30}| m \in \mathbb{Z}, 0 \leq m < 60 \}$, defining its direction of motion and receives a 
fixed negative reward $R_t$. The agent receives an additional positive reward of $100|R_t|$ and a very large penalty amounting to $R_t$ times the maximum number of steps available for each episode in case of reaching the target region or hitting the environment boundaries. Hence the optimal 
navigation strategy corresponds to 
reaching the target by taking the least number of actions necessary, without hitting the boundaries.\\
Importantly, in contrast to most previous studies, our approach neither requires any reward shaping \cite{ng1999policy, grzes2008plan} nor any heuristics \cite{ferguson2005guide, keselman2018reinforcement} and the {active particles} are able to develop the optimal strategy only through pure exploration.\\
The main component of our approach is the RL algorithm itself. It became evident during our tests, that, on-policy methods \cite{sutton2018reinforcement} provide more robustness in convergence toward the global optimal solution, {as compared to off-policy methods such as Q-learning}; hence a policy gradient method was selected (see SI for details). 
This method allows learning a 
parametrized policy $\pi_{\theta}$ for choosing actions by 
optimizing the expected return $E[J(\theta)]= E \big[ \sum_{t=0}^{T} R_t| \pi_{\theta} \big]$   \cite{sutton1999policy,peters2008reinforcement,duan2016benchmarking}. This goal is achieved by updating the set of parameters of the policy $\theta$, which in our case corresponds to the parameters {(weights)} of the policy network. \\The update is performed 
in the direction of the gradient of the expected return, i.e. such that the expected return is enhanced. 
The gradient of total expected return, averaged across $K$ trajectories, can be approximated as \cite{sutton1999policy,peters2008reinforcement,duan2016benchmarking}:

\begin{eqnarray}
        E[\nabla_{\theta} J(\pi_\theta)] \approx \frac{1}{K} \sum_{i=1}^{K} \sum_{t=0}^{T_i} \nabla_\theta log \pi_{\theta}(a_{it}|s_{it}) \sum_{t'=t}^{T_{i}} R_{t'}
    \label{eq2}
\end{eqnarray}

\noindent where $R_{t'}$ denotes the reward value at time step $t'$, achieved by policy $\pi_{\theta}$, and $T_i$, denotes the overall
lengths (number of steps) of the i-th trajectory, which can be written in state-action space as
$\xi_{i} = \{ (s_{i0},a_{i0}), (s_{i1},a_{i1}), ..., (s_{iT_{i}},a_{iT_{i}}) \}$. 
We compute these gradients via the back-propagation algorithm in the training process of the policy network. It can be seen from Eq.~(\ref{eq2}), that as the policy network improves to achieve higher total returns, the gradient of the expected return starts to diminish, until eventually the agent converges to the final policy. 
In the following we choose the advantageous actor critic method (A2C) as a specific example of a policy gradient method. 
This method involves, besides the policy network, also a critic network which assignes a value to each state which corresponds to the expected temporal distance to the target and 
is used to guide the updating of the parameters (weights) of the policy network in the following way: 
The {total return} $\sum_{t'=t}^{T_{i}} R_{t'}$ is replaced with a more well-behaved term known as the advantage function \cite{sutton1999policy,mnih2016asynchronous, schulman2015high}, which together with the critic network, essentially 
rates the possible actions by 
evaluating the benefit of choosing a specific action in comparison to the average action for a given state. {In our approach}, we define the advantage {function} as follows \cite{schulman2015high}: 
\begin{eqnarray}
    {A}^{\pi_\theta w} (s_{it},a_{it}) =  Q^{\pi_\theta}(s_{it},a_{it}) - {V}^{\pi_\theta w}
    (s_{it}) \label{advant}
\end{eqnarray}

\begin{eqnarray}
    {Q}^{\pi_\theta} (s_{it},a_{it})=  \sum_{\mu=0}^{T_{i}-{t}} \lambda^\mu R(s_{it+\mu},a_{it+\mu})
\end{eqnarray}

\noindent where $V^{\pi_\theta w}$ {is the critic network with parameters (weights) $w$, which estimates the value of a given state}, and $Q^{\pi_\theta}(s_{it},a_{it})$ {is the state-action value function under policy} ${\pi_\theta}$ and discount factor $\lambda$, which determines the expected reward when choosing a specific action.\\
We train the two networks which are involved in the A2C algorithm (policy and critic network) simultaneously based on the trajectories from multiple past episodes. 
At each training round the critic network is updated based on the 
value (discounted total return computed based on the present policy) of each state and is used to 
judge the performance of previous episodes through the advantage term, which influences (or guides) the update of the policy network through Eq. (\ref{advant}). 

\noindent Note that the Deep RL approach is able to manifest a surprisingly high level of generalization potential. Remarkably, all of the results presented in this study, across different setups with various force fields, are achieved with policy networks containing only two hidden layers and critic networks with one hidden layer. This feature vastly decreases the computational cost of training our model (see SI for details).

\section*{Results}

To demonstrate the power of the developed approach, we now consider various environments of increasing complexity, starting with cases for which exact analytical solutions are available \cite{liebchen2019optimal}.\\
As a first application of our method, we explore an active particle in a linear force (or flow) field $\vec f = (kx, 0)$. Despite its apparent simplicity, for some combinations of starting and end points, this problem 
can not be straightforwardly obtained by ``optimistic'' applications of classical path planning algorithms like  
Dijkstra or A* \cite{rao2009large} (see SI).

\begin{figure}[t]
    \includegraphics[scale = 0.119]{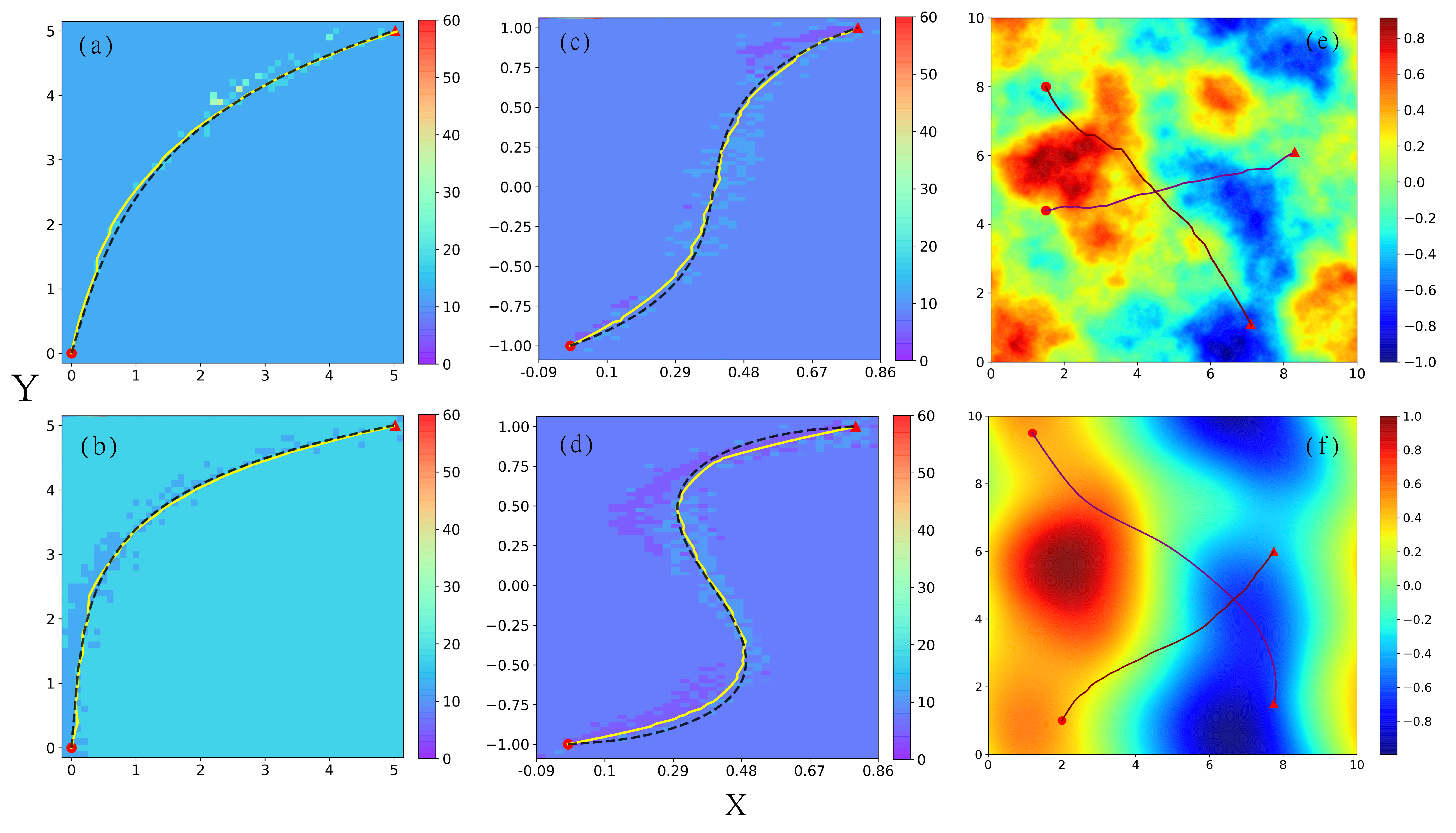}
    \caption{ Learned asymptotically optimal trajectories. (a-d) The trained active particle replicates theoretical results from \cite{liebchen2019optimal} (dashed curves), for a linear force/flow field $\vec f=(k x,0)$ with (a) $k=0.6$, (b) $k=1.0$, and for shear flow $\vec f=(k[1-y^2],0)$ with (c) $k=-0.5$, (d) $k=-0.8$. Starting and target points are shown with circles and triangles and background colors show the learned policy map. Panels (e,f) show generalizations to Gaussian random potentials with $\alpha=2$ (e) and $\alpha=4$ (f) with background colors showing the value of the Gaussian random potential. $\vec f$ is measured in units of $v_0$ (such that $v_0=1$) and computational parameters are provided in table B.1 in the SI.}
    \label{fig:GRFS}
\end{figure}

\noindent In contrast, by applying our RL-based method, we find that after training the model (see SI for details), the agent is able to achieve asymptotically optimal solutions (Figs. 3a,b). That is, any remaining deviations due to the discretization of the problem can be systematically reduced by choosing an even finer discretization. 
Notably, the algorithm finds the optimal path even for $k=1.0, \vec{r}_{start}=(0,0)$ and $\vec{r}_{target}=(5,5)$ (Fig. 3b) where the drift due to the external field dominates the self-propulsion in a large portion of space
such that
choosing unsuitable actions 
at early times can make it impossible for the agent to reach the target even after very long times.
\\{Let us now focus on a somewhat more complex environment}, defined by $\vec f = (k(1-y^2),0)$, representing e.g. shear flow in a pipe \cite{PhysRevE.84.031105}. Here the optimal path is S-shaped for sufficiently large $k$ (dashed lines in Fig. 3c and 3d).
From the RL point of view, such trajectories are rather unapparent, because achieving this symmetric S-shaped path, in the presence of the strong flow field, would require the agent to employ a strategy with intricate combination of actions. Accordingly it is not surprising that simple RL algorithms such as tabular Q-learning tend to fail finding the global optimal path or at least require unsystematic hyperparameter fine-tuning for each considered k-value (see SI for more details).
Remarkably, also for this setup, our RL agent is able to learn the asymptotically optimal path for any $k$ value. That is, also here any remaining deviations from the optimal path can be systematically reduced by choosing a finer discretization for state and action spaces. 
Notice however, that for a given discretization 
\textit{degenerate trajectories} may occur which have very similar (or even identical) temporal costs but differ slightly in shape (see SI for details). 
\\Let us now exploit the very close agreement between theoretically calculated optimal paths and the learned ones, to explore truly complex environments for which optimal trajectories cannot be analytically calculated. For this purpose, we create a Gaussian random potential (GRF) 
\cite{Pen_1997,Bertschinger_2001}
$U(\vec r)$ with a power spectrum of  $<\tilde U(k) \tilde U(-k)> \hspace{0.1cm} \propto k^{-\alpha}$ and determine $\vec{f} = - \vec{\nabla}U$ (see SI for more details).

\noindent Here we consider the cases of $\alpha = 2\,$ and $\,4$ for two different combinations of starting and end points (see Fig. 3e,f). 
Notably for $\alpha = 4$ the learned solutions are comparatively smooth curves, while for $\alpha=2$ the learned paths involve comparatively sharp turns, reflecting the structure of the underlying potential islands. 
In all shown cases, we find that the travelling time of the learned optimal trajectories is shorter than the travelling time when following a straight line.

\section*{Conclusions}

We have developed an end-to-end deep reinforcement learning approach which creates
asymptotically optimal solutions for complex navigation
problems which are not straightforwardly attainable with existing
standard methods.
This approach complements the broad literature on methods
to find the shortest path and opens a route towards a universal path
planner for microswimmers and larger self-driven agents which are subject to
continuously varying force or flow fields.

\noindent To achieve this, our method can be generalized in many
ways, e.g to find globally optimal paths with respect to
fuel consumption or dissipated power, or after generalization to continuous state-action-spaces, to account for
microswimmer-specific ingredients such as hydrodynamic
interactions with obstacles and fluctuations.\\
Potential applications of the presented method range from testing future
theoretical developments e.g. regarding the optimal path of active particles in the presence of fluctuations to the programming of
nano- and microscale robots for targeted drug and gene
delivery. On larger scales our approach could also be used to test if biological agents like turtles or fish manage to find the globally optimal path (based on a comparison of their trajectories with the learned results) or, possibly, even for route planning of macroscopic vehicles like cleaning-robots or
spacecrafts where it could provide an alternative to optimization methods
based on nonlinear programming or meta-heuristics \cite{shirazi2018spacecraft}.

\section*{Acknowledgments}
The authors would like to thank Prof. Dr. Hartmut Löwen for useful discussions.

\section*{Data availability}
The data and the code supporting the findings of this study are available from the authors upon
reasonable request.

\section*{References}
\bibliographystyle{iopart-num.bst}
\bibliography{Master}

\end{document}


\preprint{APS/123-QED}

\widetext
\begin{center}
\text{\small Supplemental Materials for}\\
\textbf{ Reinforcement learning of optimal active particle navigation}\\ \vspace{0.4cm}
\text{Mahdi Nasiri and Benno Liebchen$^*$}\\
\vspace{0.3cm}
\textit{\small Institute of Condensed Matter Physics, Technische Universität Darmstadt, D-64289 Darmstadt, Germany}
\end{center}

\email{benno.liebchen@pkm.tu-darmstadt.de}

\setcounter{equation}{0}
\setcounter{figure}{0}
\setcounter{table}{0}
\setcounter{page}{1}
\makeatletter
\renewcommand{\theequation}{S\arabic{equation}}
\renewcommand{\thefigure}{S\arabic{figure}}
\renewcommand{\bibnumfmt}[1]{[S#1]}
\renewcommand{\citenumfont}[1]{S#1}

\subsection{A2C algorithm}
In this section we will briefly review our implementation of the A2C algorithm. Just like most other policy gradient based methods, the A2C algorithm consists of two separate neural networks known as the policy (actor) and the value (critic) networks. The final decision making is based on the output of the policy net. 
In Algorithm \ref{Algo1}, a pseudo-code of the algorithm is provided. 

\begin{algorithm}[H]
  \caption{Advantageous Actor-critic (A2C)}
  \label{EPSA}
   \begin{algorithmic}[1]
   \State Define the initial policy $\theta_0$ and value $w_0$ network parameters
   \State \textbf{for} $k= 0,1,2,..., K$: 
   \State \hspace{0.5cm} Generate a set of trajectories $M_k = \{\xi_i\, | \, i \in \mathbb{N}\}$, using the policy, $\pi(\theta_k)$.
   \State \hspace{0.5cm} Compute the corresponding state-action values of each trajectory, $Q^{\pi_\theta}(\xi_i)$ based on Eq. (4).
   \State \hspace{0.5cm} Compute the advantage estimates, $\hat{A}_t (\xi_i)$ using the value network $V^{w_k}(\xi_i)$ based on Eq. (3).
   \State \hspace{0.5cm} Estimate the policy gradient:
   \begin{equation*}
        \nabla \hat{J}(\theta_k) = \frac{1}{|M_k|} \sum_{i \in M_k} \, \sum_{t=0}^{T_i} \nabla \,\text{log}\, \pi_{\theta_k}(a_t|s_t) \hat{A}_{t}.
   \end{equation*}
   \State \hspace{0.5cm} Update the policy network via Adam optimization.
   \State \hspace{0.5cm} Compute the loss on the value network:
   \begin{equation*}
       \hat{\mathcal{L}} = \frac{1}{|M_k|} \sum_{i \in M_k} \, \frac{1}{T_i} \sum_{t=0}^{T_i} \bigg (V^{w_k}(s_t) - Q^{\pi_{\theta_{k}}}(s_t, a_t) \bigg )^2.
   \end{equation*}
   \State \hspace{0.5cm} Update the value network via Adam optimization.
   
   \State \textbf{end}
   
   \end{algorithmic}
   \label{Algo1}
\end{algorithm}
As it can be seen from this pseudocode, at each step, both networks are updated using the aggregated data of $|M_k| = 10$ different trajectories created based on the current state of the policy network. 
We use an actor network with two hidden layers with 250 and 50 neurons respectively and an output layer providing the 
probability of each action in the form of a softmax function. The critic, on the other side, has only one hidden layer with $N$ neurons the value of which 
corresponds to the dimensions of the state space. We use learning rates of $10^{-4}$ and $2 \times 10^{-4}$ for the Adam optimizer \cite{kingma2014adam} in policy and value networks respectively. The training of the neural networks are done with the help of the Pytorch deep learning framework \cite{NEURIPS2019_9015}. 

\blfootnote{$^*$benno.liebchen@pkm.tu-darmstadt.de}
\subsection{Training details}
Table \ref{table:1} summarizes the parameter values which we have used for the calculation of the optimal paths in this work. Here the step size is the only parameter which is chosen based on the characteristics of the environment; the step size then determines the resolution of the discrete state space. 
 In all of our setups, a fixed negative reward (penalty) of $R_t = -0.1$ is used for each actions and we use a positive reward of $R_G=10$ ($R_G=20$ in case of the Gaussian random field with $\alpha=4$) for reaching the target domain and a penalty of 
 $\text{Max \# of steps} \times R_t$ ($2\times \text{Max \# of steps} \times R_t$ for the Gaussian random field with $\alpha=4$) for hitting the boundary.
  The environments are implemented using OpenAI's  gym framework \cite{brockman2016openai}.

\begin{table}[H]
\centering
\renewcommand\thetable{B.1}
\begin{tabular}{||c c c c c||} 
\hline 
 Environment & $k$ & Resolution & Step size &  Training Period ($\times 10^3$) \\ [0.5ex] 
 \hline\hline 
 Linear Flow (I) & 0.6 & $53\times53$ & 0.1 & 260  \\  [1ex]
 Linear Flow (II) & 1.0 & $53\times53$ & 0.1 & 300  \\  [1ex]
 Shear Flow (I) & -0.5 & $86\times38$ & 0.025 & 170  \\  [1ex]
 Shear Flow (II) & -0.8 & $86\times38$ & 0.025 & 170\\  [1ex]
 GRF $\alpha=2$ (I) & 0.25 & $50\times50$ & 0.2 & 160  \\  [1ex]
GRF $\alpha=2$ (II) & 0.25 & $50\times50$ & 0.2 & 180  \\  [1ex]
GRF $\alpha=4$ (I) & 1.5 & $50\times50$ & 0.2 & 200  \\  [1ex]
GRF $\alpha=4$ (II) & 1.5 & $50\times50$ & 0.2 & 200  \\  [1ex]
 \hline
\end{tabular}
\caption{Parameter values and computational details used to train the deep learning algorithm. Training periods correspond to number of iterations (episodes).}
\label{table:1}
\end{table}

\subsection{Method to obtain the exact solutions}
For the exact (analytical) solutions of each setup, the final forms of the trajectories, $y'(x)$ and the boundary conditions, $y(x_A) = y_A$ and $y(x_B) = y_B$ are the same as in \cite{liebchen2019optimal}. We utilize a modified version of the shooting method to numerically solve these equations, in which we couple an optimization process to the ordinary shooting method that automatically updates the required constant $c_0$ in $y'(x)$, for each round of the shooting approach. For this purpose, at the first step a random small value is given to $c_0$ and after solving the equation, the last value of the achieved solution is compared with the boundary condition and the resulting difference $\Delta = y(x_B) - \hat{y}(x_B) $ is used to update the initial constant, $c_0' = c_0 + \beta \Delta $, where $\beta$ and $\hat{y}$ correspond to the learning rate and the numerically estimated solution. This process is continued until the estimated value of the boundary condition reaches a certain accuracy threshold, $\epsilon$.

\subsection{Classical path planners and Q-learning}
\begin{figure}[h]
    
    \includegraphics[scale = 0.16]{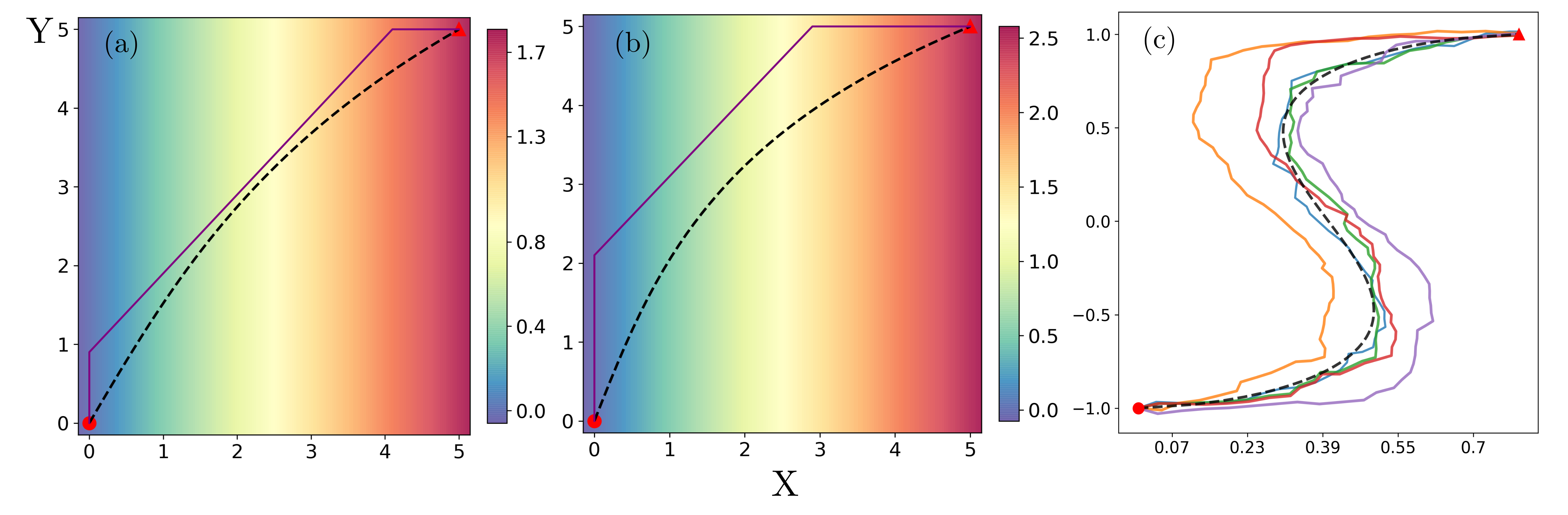}
    \renewcommand\thefigure{D.1}
    \caption{Optimal trajectory via Dijkstra, A*, and tabular Q-learning algorithms. Results achieved by applying the path planners on linear flows with (a) $k=0.35$, (b) $k=0.5$, depicted by purple line. The background color and corresponding color maps represent the field strength.(c) Examples of learned trajectories using the tabular Q-learning algorithm in the environment hosting the shear flow. Each color represents an independent run with its unique random seed. Exact analytical solutions are presented by dashed black lines for each case. }
    
    \label{fig:Dijkstra}
\end{figure}

To further illustrate the power of the developed method, we now use two types of existing approaches to determine the optimal path (to the extend possible). We first discuss the Dijkstra and the A* algorithms as examples of classic path planning methods and then turn our attention to the tabular Q-learning algorithm. 
While the Dijkstra algorithms has been mainly designed to find the shortest path, it
can be heuristically transformed into an algorithm aiming to find the fastest path
by defining a temporal cost of $t_{cell} = |\vec{\Delta X}|/|\dot{\vec{r}}_{tot}|$. Here $|\vec{\Delta X}|$ is the displacement length in a certain step (which is larger for diagonal moves than for horizontal and vertical ones) and $|\dot{\vec{r}}_{tot}|$ is the maximum total velocity possible for that displacement (which must be a positive real value).
For the implementation of the algorithm we follow \cite{rao2009large}. 
At each step we allow the agent to choose between eight equidistantly chosen possible directions. 
To illustrate the power of our deep RL approach as compared to these classic path finders, we focus on the simplest case of a linear flow or force field. The Dijkstra algorithm was able to achieve decent results for weak flow field (low $k$), closely approximating the analytical solutions (Fig \ref{fig:Dijkstra}a), however once the force field dominates over the self-propulsion, the results started significantly deviate from the exact solutions (Fig \ref{fig:Dijkstra}b).
Let us now discuss the A* algorithm. 
Here the same cost is used together with a heuristic of $h(s_t) = d_{heur}/v_{heur}$, where $d_{heur}$ is the minimum distance to the target from the current cell, and $v_{heur} = \text{max}(|\dot{\vec{r}}_{tot}|)$. As it can be seen, the trajectories predicted by the A* algorithm turn out to be identical to those of the modified form of the Dijkstra algorithm. These examples depict the limitations of classical path planners for finding shortest time trajectories, in complex environments. 
For these methods to achieve asymptotically optimal solutions, one usually has to balance between the generalization capabilities of these approaches and their accuracy by designing task specific heuristics which in return would need to be modified in case of changes to the environment or the underlying flow fields. 

Finally a tabular Q-learning agent was implemented with the same 60-dimensional action space as used for the on-policy approach which we have discussed in the main text. To assess the power of this algorithm and its limitations, we trained this agent on the environment hosting the shear flow $f(x,y) = (k(1-y^2,0)$ with $k=-0.8$. The same hybrid observation of the environment with the resolution of $43\times19$ was fed to train the model with \textit{learning rate= $0.1$}, exploration ($\epsilon-greedy$) parameter $\epsilon_0 = 0.25$,  and \textit{discount factor}=$0.9$ for $4\times 10^6$ generations, More details of this algorithm can be found in \cite{sutton2018reinforcement}. Given these specifications of the environment and the algorithm, we were able to train the model with a reward of $R_t = -1$, target reward $R_{G} = 0.5$, and boundary penalty $R_{B} = -100$. Note that even small modifications to these values or the hyperparameters turned out to drastically affect the outcome trajectories. In particular, when choosing a much larger target reward, the results did not improve.\\ As it can be seen from some representative examples of the converged results (Fig ~\ref{fig:Dijkstra} c), most of the achieved trajectories are far from the optimal solution as opposed to the performance of our on-policy agent. It is important to note that although one may be able to achieve asymptotically optimal solutions using Q-learning via brute force grid searching in the hyperparameter space,  problems arise due to the instability of the final converged results. That is for any independent training session, there is the possibility of converging to a totally different outcome. Also the discovered hyperparameter set lacks the necessary robustness with regards to changes in the environment or the underlying flow fields, and a slight modification of the flow field strength ($k$-value) will require a renewed grid search in the hyperparamter space. Note that increasing the dimensionality of the state space would decrease the degeneracy of the temporal cost of the outcome trajectories, but would not help to systematically address the instability of the outcome trajectories or the sensitivity of the results to hyperparameter definitions.  \\
One of the main reasons for this limited performance of simple off-policy methods such as tabular Q-learning could be rooted in the partial observability of the underlying state space of our hybrid setup. This characteristics of our state space is turning the path planning task into a complicated problem since in the presence of a strong flow field, small modifications in the navigation strategy would result in different outcomes for a given state and the final trajectory, which is not easily comprehensible to a simple off-policy method such as tabular Q-learning given the current coarse-grained image of the environment.

\subsection{Degenerate trajectories}
\begin{figure}[t]

    \includegraphics[scale = 0.4]{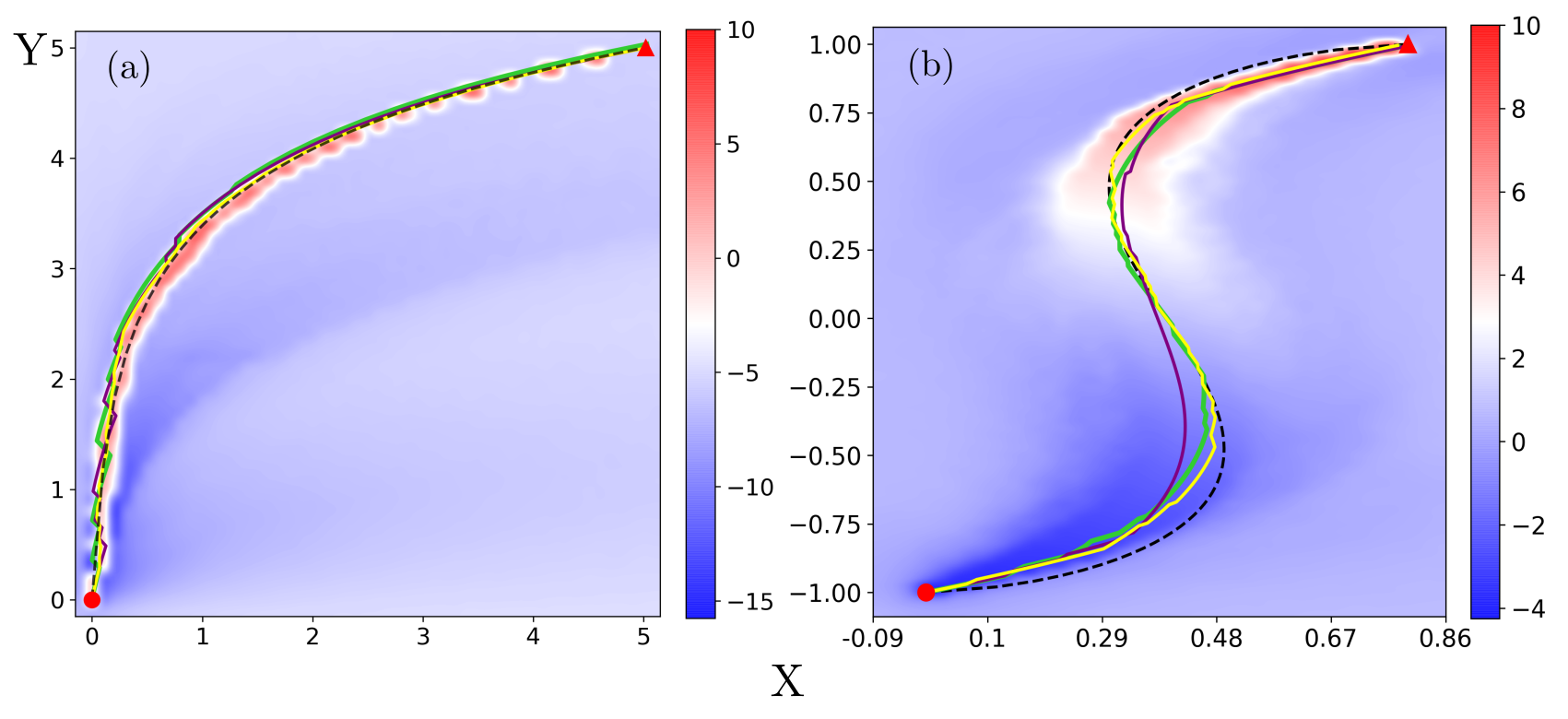}
    \renewcommand\thefigure{E.1}
    \caption{Degeneracy in the converged results. Colors (green,
purple) represent different examples of degenerate trajectories with
identical temporal length. Yellow trajectories show the average over 20 independent training sessions with various neural network initialization random seeds. Background colors show the values as
predicted by the critic network for a given discretization. Parameters and setups are as in Fig. 3
b,d respectively.}

    \label{fig:Degen}
\end{figure}

In all of our setups, at the end of the training, the RL agent does not always converge to one and the same trajectory, but it alternatingly chooses a trajectory out of a set of distinct trajectories which have identical temporal costs for a given discretization (but have different temporal length in continuum space).
This phenomena is expected due to the coupling of the intrinsic stochasticity of the training process of the neural networks \cite{PytorchRepro} and the hybrid definition of our state spaces. The effects of these characteristics on the reproducability of the results decrease as finer discretizations are chosen for the state-action spaces. It is expected, in the limit of a very fine grid all of these degenerate trajectories would converge to the same optimal path in continuum space. 
To further illustrate this matter, it is instructive to look at the final learned strategy from the critic network's point of view. As it can be seen from the value map (background colors in Fig. \ref{fig:Degen}), the values as predicted by the critic network are essentially constant in
the vicinity of the optimal path in continuum space (dashed line).

\subsection{Gaussian random field}
Our Gaussian random potentials are realized based on the power spectrum
realization method \cite{Pen_1997, GRF}. To this end, we have first created a
$500\times500$ matrix of uncorrelated Gaussian
random numbers (white noise) with values $\in [0,1]$
using \textit{Numpy}.
After Fourier transforming the result (fftpack module in \textit{SciPy})
and multiplying the result with
the square root of the power spectrum, the new values are transformed back.
The final values are rescaled to reside in the $[-1,1]$ region. After
achieving the potential field, the force fields are computed via finite
differences and a coefficient, $k$ is multiplied to rescale the
strengths of force fields. Note that for a state space with the
resolution of $50\times50$, an underlying force field of $500\times500$
is realized.

\bibliographystyle{iopart-num.bst}
\bibliography{SuppInfo}